\documentclass[12pt,preprint]{aastex}

\usepackage{amsmath}

\slugcomment{Not to appear in Nonlearned J., 45.}

\shorttitle{Singular Isothermal Quadrupole Lens}
\shortauthors{Chu et al.}

\begin{document}


\title{Analytical Solutions of Singular Isothermal Quadrupole Lens}


\author{Zhe Chu\altaffilmark{1,2}, W. P. Lin\altaffilmark{1,2,3}, and Xiaofeng Yang\altaffilmark{4} }
\altaffiltext{1}{Key Laboratory for Research in Galaxies and Cosmology, Shanghai Astronomical Observatory,
Chinese Academy of Sciences, 80 Nandan Road, Shanghai 200030, China}
\altaffiltext{2}{University of the Chinese Academy of Sciences, 19A Yuquan Road, Beijing 100049, China}
\altaffiltext{3}{Center for Astronomy and Astrophysics, Shanghai Jiao Tong University, Shanghai 200240, China}
\altaffiltext{4}{School of Astronomy and Space Science, Nanjing University, Nanjing 210093, China}
\email{chuzhe,linwp@shao.ac.cn}

\begin{abstract}

Using analytical method, we study the Singular Isothermal Quadrupole (SIQ) lens system, which is the simplest lens model that can produce four images. In this case, the radial mass distribution is in accord with the profile of the Singular Isothermal Sphere (SIS) lens, and the tangential distribution is given by adding a quadrupole on the monopole component. The basic properties of the SIQ lens have been studied in this paper, including deflection potential, deflection angle, magnification, critical curve, caustic, pseudo-caustic and transition locus. Analytical solutions of the image positions and magnifications for the source on axes are derived. As have been found, naked cusps will appear when the relative intensity $k$ of quadrupole to monopole is larger than 0.6. According to the magnification invariant theory of the SIQ lens, the sum of the signed magnifications of the four images should be equal to unity \citep{dal98}. However, if a source lies in the naked cusp, the summed magnification of the left three images is smaller than the invariant 1. With this simple lens system, we study the situations that a point source infinitely approaches a cusp or a fold. The sum of magnifications of cusp image triplet is usually not equal to 0, and it is usually positive for major cusp while negative for minor cusp. Similarly, the sum of magnifications of fold image pair is usually neither equal to 0. Nevertheless, the cusp and fold relations are still equal to 0, in that the sum values are divided by infinite absolute magnifications by definition.
\end{abstract}


\section{Introduction}

The Singular Isothermal Sphere (SIS) lens is a very important lens model in strong gravitational lenses. The mass distribution of this lens yields flat rotation curve. More practical is the Singular Isothermal Ellipsoid (SIE) lens model, derived by extending the mass distribution of the SIS lens into elliptical distribution \citep{kas93}. There are also some other generalized SIS lenses, with SIS radial profile and arbitrary tangential distribution \citep{eva01,eva03}. The Singular Isothermal Quadrupole (SIQ) lens is one of the generalized SIS lenses whose tangential distribution function is given by adding a quadrupole on the monopole component. Therefore, the SIQ lens is often regarded as SIS lens perturbed by an external quadrupole shear. In general, critical curves and caustics can be divided into tangential and radial ones. Because the singular isothermal lens does not formally have a radial critical curve, the radial caustic associated with the singularity in lens center is usually called pseudo-caustic (or cut). When a point source traverses such a pseudo-caustic, the image number changes by $\pm1$, instead of $\pm2$ for the general caustic \citep{eva98}. The tangential caustic of a smooth quadrupole lens commonly comprises four cusps and four folds. The transition loci in the lens plane are also very important for understanding the spatial distribution of multiple images. The transition loci together with critical curves are the images of caustics. For detailed information about transition loci, please refer to \citet{fin02,an06,chu13}.

There are some important magnification relations for the multiple image lenses. The magnification invariant means for some specific lens models, the sum of signed magnifications for all lensed images of a given source is a constant \citep{wit95,wit00,hun01}. It is very interesting and surprising that the invariants are independent of most of the model parameters. For the SIQ lens, the sum of the four magnifications equals unity \citep{dal98,dal01}, i.e., $\sum_{i=1}^{4}\mu_{i}=1$ (here $\mu$ is the signed magnification, same hereafter). In these previous works, the magnification invariant was calculated through polynomial coefficients to avoid deriving each magnification of the images. The cusp and fold relations are {\em local} magnification relations compared with the magnification invariant \citep{pet10}. It has been found that, if a point source approaches the cusp infinitely from the inner side of the tangential caustic, the three close images will have a relation with $R_{cusp}=\sum_{i=1}^{3}\mu_{i}/\sum_{i=1} ^{3}|\mu_{i}|=0$  \citep{bla86,sch92,mao98}. A similar magnification relation holds when the source lies near a fold caustic, however the prerequisite is that the source position should not be very close to the cusp. If the source infinitely approaches the fold line, the fold image pair will also have a relation with $R_{fold}=\sum_{i=1}^{2}\mu_{i}/\sum_{i=1}^{2}|\mu_{i}|=0$ \citep{bla86,gol10}. In this Letter, we would like to present analytical solutions and some interesting results for the SIQ lens.

This Letter is arranged as follows. In Section 2, some basic properties of the SIQ lens are presented, including deflection potential, deflection angle, magnification, critical curve, caustic, pseudo-caustic and transition locus. In Section 3, the analytical solutions of the image positions and magnifications for point source on the axes are calculated. Then the general case with arbitrary source position is extended. In Section 4, the lost image and the magnification relations are investigated, including magnification invariant, cusp and fold relations. Finally, conclusions and discussion will be given in Section 5.

\section{Basic Properties of the SIQ Lens}

Figure 1 shows the main curves of the SIQ lens in the source plane and lens plane. If a source lies in the astroid caustic, there are usually four images produced. They are distributed in the four image regions, which are divided by the critical curve and transition loci. The sign of the images 1 and 3 is positive, while it is negative for the images 2 and 4. The convergence (surface mass density in units of the critical surface mass density) of the SIQ lens is given by adding a quadrupole component on the SIS lens
\begin{equation}
\kappa(\boldsymbol{\theta})=\frac{\theta_{E}}{2\theta}(1+k\cos 2\phi),
\end{equation}
where $(\theta, \phi)$ are the polar coordinates and $\theta_{E}$ is the Einstein radius of the SIS lens. Here $k$ ($0\leqslant k\leqslant1$) is the intensity of the quadrupole relative to the monopole.

The deflection potential depends on the two dimensional Poisson equation $\nabla^2\psi(\boldsymbol{\theta})=2\kappa(\boldsymbol{\theta})$. By solving this differential equation in polar coordinate, one can derive the deflection potential
\begin{equation}
\psi(\boldsymbol{\theta})=\theta_{E}\theta(1-\frac{1}{3}k\cos 2\phi) .
\end{equation}
The scaled deflection angle $\boldsymbol{\alpha}=\nabla\psi(\boldsymbol{\theta})$ is the first derivation of deflection potential. Its radial and tangential components are
\begin{equation}
\alpha_{rad}=\theta_{E}-\frac{1}{3}\theta_{E}k \cos 2\phi ,
\end{equation}
\begin{equation}
\alpha_{tan}=\frac{2}{3}\theta_{E}k \sin 2\phi .
\end{equation}

The shear can be calculated through the deflection potential $\psi$. For this lens model, $\gamma=\kappa$ \citep{chu13}. Consequently, the magnification in the lens plane is
\begin{equation}
\mu=\frac{1}{(1-\kappa)^2-\gamma^2}=\frac{\theta}{\theta-\theta_{E}-\theta_{E}k\cos 2\phi} ,
\end{equation}
and therefore, the critical curve where $\mu$ is infinite can be derived as
\begin{equation}
\theta=\theta_{E}+\theta_{E}k \cos 2\phi .
\end{equation}

The lens equation $\boldsymbol{\beta}=\boldsymbol{\theta}-\boldsymbol{\alpha}$, describing the transformation between the lens plane $(\theta, \phi)$ and the source plane $(\beta, \varphi)$, can be written in the polar coordinate as \citep{chu13}
\begin{equation}
\beta^2=(\theta-\alpha_{rad})^2+\alpha_{tan}^2 ,
\end{equation}
\begin{equation}
\tan(\phi-\varphi)=\frac{\alpha_{tan}}{\theta-\alpha_{rad}} .
\end{equation}

The major and minor cusps correspond to the points with $\cos 2\phi=\pm1$ on the critical curve, so we can obtain $\alpha_{tan}=0$. Then from Equations (3), (6) and (7), one can derive the distances from the major or minor cusps to the center of the source plane. Both of them are exactly
\begin{equation}
\beta_{cusp}=\frac{4}{3}\theta_{E}k .
\end{equation}

Putting Equations (3)-(4) and (6) into Equation (8), and using the trigonometric functions, the phase relation between the corresponding points on critical curve and caustic can be obtained
\begin{equation}
\tan^3\phi=-\tan \varphi .
\end{equation}
Inserting Equations (3)-(4) and (6) into Equation (7), $\beta$ of caustic can be derived as
\begin{displaymath}
\beta=\frac{1}{2}\beta_{cusp}\sqrt{1+3\cos^2 2\phi} .
\end{displaymath}
Then, by substituting $\phi$ with $\varphi$ through Equation (10), the caustic is given as
\begin{equation}
\beta=\beta_{cusp}\frac{\sqrt{\tan^{4/3}\varphi-\tan^{2/3}\varphi+1}}{\tan^{2/3}\varphi+1} .
\end{equation}
It can be easily found that the caustic is symmetric respective to the axes $\varphi=\pm\pi/4$.

Because the pseudo-caustic corresponds to the singular point in the lens center, its expression can be easily derived by setting $\theta=0$ in the lens equations (7)-(8). The transition locus can be calculated by putting Equation (11) into Equations (7)-(8) again. The pseudo-caustic and transition loci can be obtained in the form of parametric equations. However their expressions are very complex.

\section{Positions and Magnifications of the Multiple Images}

Given a position $(\beta, \varphi)$ in the source plane, one can calculate the corresponding image positions $(\theta, \phi)$ in the lens plane through lens equations. It is relatively easier to calculate $\phi$ first. Therefore we eliminate $\theta$ through Equations (7) and (8) to obtain
\begin{displaymath}
\left(\frac{\tan\phi-\tan\varphi}{1+\tan\phi \tan\varphi}\right)^2=\frac{\alpha_{tan}^2}{\beta^2-\alpha_{tan}^2} ,
\end{displaymath}
where the tangential deflection angle can also be written as
\begin{displaymath}
\alpha_{tan}=\frac{\beta_{cusp}\tan\phi}{1+\tan^{2}\phi} .
\end{displaymath}
Thus, the expression of lens equation is changed with only one unknown quantity $\tan\phi$. Then we set $\eta=\tan\phi$ and $\lambda=\tan\varphi$ to get a simple expression
\begin{equation}
\left(\frac{\eta-\lambda}{1+\eta\lambda}\right)^2=\frac{\beta_{cusp}^2\eta^{2}}{\beta^2(1+\eta^{2})^2-\beta_{cusp}^2\eta^{2}} .
\end{equation}

\subsection{Source on the Major Axis}

For the positive direction of the major axis in the source plane, it has $\varphi=0$ and $\lambda=0$. The left-hand side of the equal sign in Equation (12) turns into $\eta^2$. Apparently, $\eta=0$ satisfies the equation, so we have two solutions $\phi_{2}=\pi$ and $\phi_{4}=0$. In addition, after eliminating $\lambda$, Equation (12) can be organized into
\begin{equation}
\beta^2\eta^4+(2\beta^2-\beta_{cusp}^2)\eta^2+\beta^2-\beta_{cusp}^2=0 .
\end{equation}
Disregarding the unphysical solution $\eta^2=-1$, we have
\begin{equation}
\eta^2=\frac{\beta_{cusp}^2-\beta^2}{\beta^2} .
\end{equation}
Thus, $\tan\phi_{1,3}=\pm\sqrt{\beta_{cusp}^2-\beta^2}/\beta$. The two solutions are invalid when $\beta>\beta_{cusp}$. Substituting the four solutions of $\phi$ into Equation (7) or (8), the radii $\theta$ will be
\begin{equation}
\theta_{2}=\theta_{E}-\frac{1}{4}\beta_{cusp}-\beta ,
\end{equation}
\begin{equation}
\theta_{4}=\theta_{E}-\frac{1}{4}\beta_{cusp}+\beta ,
\end{equation}
\begin{equation}
\theta_{1,3}=\theta_{E}+\frac{1}{4}\beta_{cusp}+\frac{\beta^2}{2\beta_{cusp}} .
\end{equation}
One can find that $\theta_{1,3}$ are always larger than $\theta_{2,4}$. Putting the four solutions of $\phi$ and $\theta$ into Equation (5), and substituting $k$ using Equation (9), the magnifications are derived
\begin{equation}
\mu_2=-\frac{4\theta_E-\beta_{cusp}-4\beta}{4\beta_{cusp}+4\beta} ,
\end{equation}
\begin{equation}
\mu_4=-\frac{4\theta_E-\beta_{cusp}+4\beta}{4\beta_{cusp}-4\beta} ,
\end{equation}
\begin{equation}
\mu_{1,3}=\frac{4\theta_E\beta_{cusp}+\beta_{cusp}^2+2\beta^2}{4\beta_{cusp}^2-4\beta^2} .
\end{equation}
Finally, one can obtain the exact relation $\mu_1+\mu_2+\mu_3+\mu_4=1$.

\subsection{Source on the Minor Axis}

For the positive direction of the minor axis in the source plane, it has $\varphi=\pi/2$ and $\lambda=\infty$. The left-hand side of the equal sign in Equation (12) turns into $1/\eta^2$. Apparently, $\eta=\infty$ is the solution, in other words $\phi_{1}=\pi/2, \phi_{3}=-\pi/2$. In addition, after eliminating $\lambda$, Equation (12) can be organized into
\begin{equation}
(\beta^2-\beta_{cusp}^2)\eta^4+(2\beta^2-\beta_{cusp}^2)\eta^2+\beta^2=0 .
\end{equation}
Disregarding the unphysical solution $\eta^2=-1$, we have
\begin{equation}
\eta^2=\frac{\beta^2}{\beta_{cusp}^2-\beta^2} .
\end{equation}
Thus, $\tan\phi_{2,4}=\pm\beta/\sqrt{\beta_{cusp}^2-\beta^2}$. The two solutions are also invalid when $\beta>\beta_{cusp}$. Putting the four solutions of $\phi$ into Equation (7) or (8), the radii $\theta$ will be
\begin{equation}
\theta_{1}=\theta_{E}+\frac{1}{4}\beta_{cusp}+\beta ,
\end{equation}
\begin{equation}
\theta_{3}=\theta_{E}+\frac{1}{4}\beta_{cusp}-\beta ,
\end{equation}
\begin{equation}
\theta_{2,4}=\theta_{E}-\frac{1}{4}\beta_{cusp}-\frac{\beta^2}{2\beta_{cusp}} .
\end{equation}
As before, $\theta_{1,3}$ are always larger than $\theta_{2,4}$. Putting the four solutions of $\phi$ and $\theta$ into Equation (5), and substituting $k$ using Equation (9), the magnifications are derived
\begin{equation}
\mu_1=\frac{4\theta_E+\beta_{cusp}+4\beta}{4\beta_{cusp}+4\beta} ,
\end{equation}
\begin{equation}
\mu_3=\frac{4\theta_E+\beta_{cusp}-4\beta}{4\beta_{cusp}-4\beta} ,
\end{equation}
\begin{equation}
\mu_{2,4}=-\frac{4\theta_E\beta_{cusp}-\beta_{cusp}^2-2\beta^2}{4\beta_{cusp}^2-4\beta^2} .
\end{equation}
Again, one will obtain the exact relation $\mu_1+\mu_2+\mu_3+\mu_4=1$.

\subsection{Source Position for the General Case}

For the general case that a point source lies at an arbitrary position, Equation (12) can be reorganised into
\begin{equation}
(\eta^2+1)[\eta^4-2\lambda\eta^3-\gamma^2(1+\lambda^2)\eta^2-2\lambda\eta+\lambda^2]=0 ,
\end{equation}
where $\gamma=\sqrt{\beta_{cusp}^2-\beta^2}/\beta$. The redundant solution $\eta^2=-1$ also appears as before. This quartic equation looks complex but in principle can be solved analytically. It has four analytical solutions, and each of them is function of $\lambda(\varphi)$ and $\gamma(\beta)$. As has been derived above, when $\lambda=0$ or $\lambda=\infty$, we have the solutions on the axes that $\eta=\pm\gamma$ or $\eta=\pm1/\gamma$. \citet{wol12} have also studied angular distributions of the four images of SIQ lens intensively. If $\varphi=\pi/4$, and $\lambda=1$, then the quartic equation in Equation (29) turns into
\begin{equation}
\eta^4-2\eta^3-2\gamma^2\eta^2-2\eta+1=0 .
\end{equation}
Although the form is very simple, the analytical solutions are still complex.

Alternatively the solutions of Equation (29) can be calculated numerically. Then, using Equation (8) again, one can obtain the result of radii
\begin{equation}
\theta=\theta_{E}+\frac{\beta_{cusp}}{4}\frac{\eta^3+3\lambda\eta^2+3\eta+\lambda}{\eta^3-\lambda\eta^2+\eta-\lambda} .
\end{equation}
Furthermore, using Equations (5) and (31), one can calculate the magnifications
\begin{equation}
\mu=\frac{(4\theta_E+\beta_{cusp})\eta^3-(4\theta_E-3\beta_{cusp})\lambda\eta^2+(4\theta_E+3\beta_{cusp})\eta-(4\theta_E-\beta_{cusp})\lambda}{4\beta_{cusp}(\eta^3+\lambda)} .
\end{equation}

\section{The Lost Image and Magnification Relations}

The naked cusps in the quadrupole lens often make problems more complex. However, they are also very intriguing. By setting $\phi=0$ ($\phi=\pi/2$) and $\theta=0$ in Equation (7), one can find that, on the major (minor) axis, the distance from the pseudo-caustic to the center of the source plane is
\begin{equation}
\beta_{pseu}=\theta_{E}\mp\frac{1}{3}\theta_{E}k .
\end{equation}
Therefore, the distance on the major axis is smaller than that on the minor axis. Comparing with $\beta_{cusp}$ in Equation (9), one can find that, on the minor axis, $\beta_{pseu}$ is always larger than or equal to $\beta_{cusp}$, which means the naked cusp never appear near the minor cusp. However, on the major axis, $\beta_{pseu}$ can be smaller than $\beta_{cusp}$ as long as $k>0.6$, which means that the naked cusp can appear near the major cusp.

It is very interesting to study the lost image when a source locates in the naked cusp. In this case, $k>0.6$ and $\beta_{pseu}<\beta<\beta_{cusp}$, the value $\theta_2$ of Equation (15) will be smaller than 0, while the value $\mu_2$ of Equation (18) will be larger than 0. Figure 2 shows the magnifications of different images when $k=0.6$. In the left panel, the intersection point of the blue curve and dashed line $\mu=0$ sits just on the cusp. If $k<0.6$, the intersection point will move to the right. If $k>0.6$, the intersection point will move to the left in stead. Therefore, when naked cusp appears, the magnification $\mu_{2}$ of the lost image has a positive sign.

Naked cusp will never appear on the minor axis. However, as shown in Figure 1, if a source continue moves outward along the axis, it will also traverse the pseudo-caustic and result in an image disappearing in lens center. After the source traversing the pseudo-caustic on the minor axis, the value $\theta_3$ of Equations (24) will be smaller than 0, while the value $\mu_3$ of Equations (27) will be larger than 0. From Equation (9) and (33), one can find on the minor axis, when $k=0.6$, $\beta_{pseu}=1.5\beta_{cusp}$, which is also shown by intersection point of the red curve and dashed line $\mu=0$ in the right panel of Figure 2.

In Section 3, we have proved that the relation $\sum_{i=1}^{4}\mu_{i}=1$ is always valid on the axes, even for $\beta>\beta_{cusp}$. However, this equations include magnifications of the lost images. As long as a source lies outside of the astroid caustic or the pseudo-caustic, the number of the images will be less than four, so the sum of magnifications of the total images is no longer equal to 1. When a source locates in the naked cusp, there are three images left, including two with positive parity and one with negative parity, and the sum of magnifications will be smaller than the invariant 1.

According to the magnification invariant, when a source approaches the cusp infinitely, the summed magnification of the triple infinite images is usually not equal to 0, except that the finite magnification of the fourth image equals 1. The sum of magnifications of the triple images for major and minor cusps can be written as
\begin{equation}
\sum_{i=1}^{3}\mu_{i}=\frac{\pm4\theta_E+3\beta_{cusp}}{4\beta_{cusp}+4\beta} .
\end{equation}
If the source locates exactly on the cusp, Equation (34) turns into $\frac{3}{8}(1\pm\frac{1}{k})$. For the major cusp, it is apparently larger than 0, while for the minor cusp, it is smaller than 0 and can be equal to 0 only when $k=1$. If $k$ is very small, the summed magnification can be a very large value. Nevertheless, after they are divided by the sum of absolute magnifications, they will become as smaller as 0 and the cusp relation is still valid with $R_{cusp}=0$.

The sum of magnifications of fold image pair can be derived from the summed value of the two images which have finite magnifications. It is similar that the sum of magnifications of fold image pair is usually not equal to 0, except that the summed magnification of two finite images is equal to 1. Obviously, it is hardly valid in ordinary situations. The summed magnification of the infinite fold image pair is finite, as long as the point source on the fold is not infinitely close to the cusp. For the SIQ lens, the summed magnification of fold image pair changes continuously along the fold line. In general, the nearer from the major cusp, the smaller the summed magnification is, while the nearer from the minor cusp, the larger the magnification. After the sum of magnifications is divided by the sum of absolute magnifications, it will turn into 0 and the fold relation is still valid with $R_{fold}=0$. Rather, in the case that a source on the fold is infinitely approaching the cusp, the value of the fold relations will be $R_{fold}=\mp\frac{1}{3}$ for the major and minor cusps, respectively \citep{kee05,aaz06}.

\section{Conclusions and Discussion}

In this work, we mainly focus on the SIQ lens model, which can produce the simplest critical curve and caustic among all of the four-image lenses. Four-image lens systems are very important and are very common in the observations of lensed quasars \citep{rus01}. Using analytical methods, we study some basic properties of the SIQ lens in the polar coordinate, including deflection potential, deflection angle, magnification, critical curve, caustic, pseudo-caustic and transition locus. Analytical solutions of image positions and magnifications for sources locating on axes are derived. Consequently, we verify the magnification invariant theory for the source on the axes of the SIQ lens, that the sum of magnifications of the four images is equal to 1. If the relative intensity $k$ of the quadrupole to the monopole is larger than 0.6, naked cusps will appear. It is found that, if the source locates in the naked cusp, there are only two positive and one negative images left, and the sum of magnifications of the triplet will be smaller than the invariant 1.

In previous works, when the positions or magnifications of the cusp or fold images are calculated, the higher order terms of Fermat potential are usually ignored \citep{sch92,zak95,con08}. In the popular views, for a point source infinitely approaching a cusp or a fold, the summed magnifications of cusp triplet or fold pair are considered to be 0, which are independent of lens models \citep{zak95,aaz09,wer09}. The summed magnifications are higher order infinitesimals compared to the magnifications of the cusp or fold images. Nevertheless, when the magnifications of these images are infinite, the higher order terms of Fermat potential can bring a significant effect. Through this simple lens model, we found that, when the point source infinitely approaches the cusp from the inner side, the summed magnification of the triple images is usually not equal to 0. For the major cusp of the SIQ lens, the summed magnification is larger than 0, while for the minor cusp, it is smaller than 0 and can be equal to 0 only when $k=1$. Similarly, if the point source approaches the fold infinitely, the summed magnification of the fold double images is usually not equal to 0 either. In general, for the fold image pair of a smooth lens, the nearer from the major cusp, the smaller the summed magnification is, while the nearer from the minor cusp, the larger the magnification. However, after the summed values are divided by the sum of absolute magnifications, they are equal to 0, so that the cusp and fold relations are still valid. These results derived from the SIQ lens may be important for explaining the anomalous flux ratio problem of the lensed quasars.



\acknowledgments

W.P.L. acknowledges supports by NSFC projects (No. 10873027, 11121062, 11233005) and the Knowledge Innovation Program of the Chinese Academy of Sciences (grant KJCX2-YW-T05).

\clearpage

\begin{figure}
\epsscale{.80}
\plotone{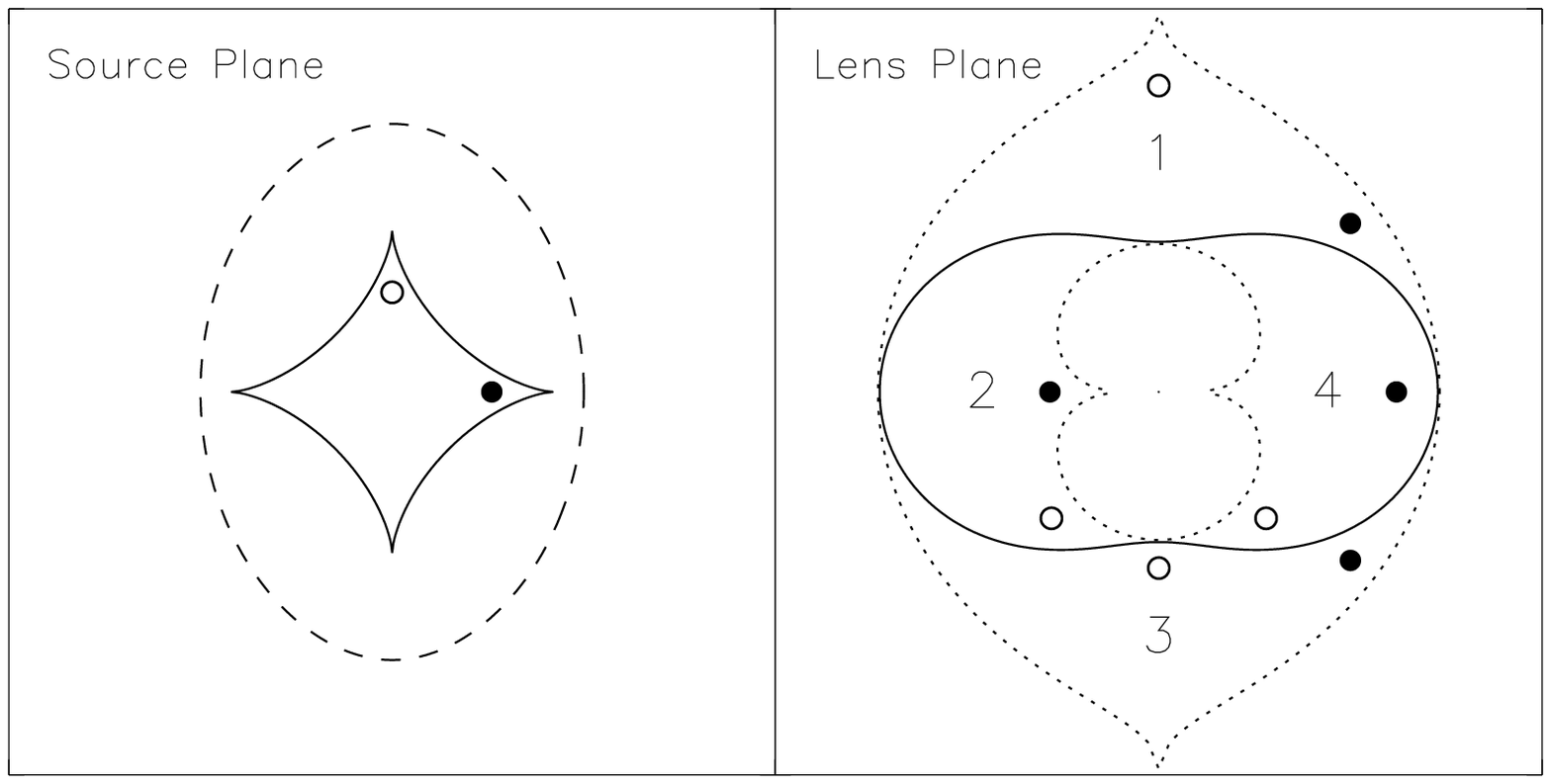}
\caption{Left panel: astroid caustic (solid curve) and pseudo-caustic (dashed curve), with solid and open circular sources near the major and minor cusps, respectively; Right panel: critical curve (solid curve) and inner and outer transition loci (doted curve), as well as the images of the two circular sources. \label{fig1}}
\end{figure}

\begin{figure}
\epsscale{.80}
\plotone{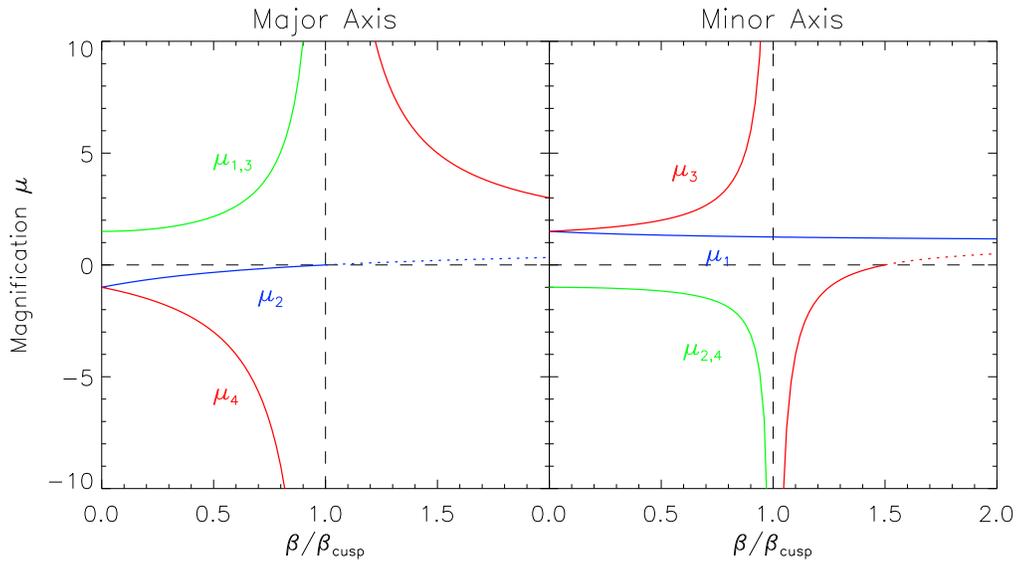}
\caption{Left panel and right panel show the image magnifications for sources locating on the major and minor axes, respectively. The intensity of the quadrupole relative to the monopole is set to be $k=0.6$. The dotted curve represents the magnification of the image lost in lens center. \label{fig2}}
\end{figure}




\clearpage



\begin{thebibliography}{}

\bibitem[Aazami \& Natarajan(2006)]{aaz06} Aazami, A. B., \& Natarajan, P. 2006, \mnras, 372, 1692
\bibitem[Aazami \& Petters(2009)]{aaz09} Aazami, A. B., \& Petters, A. O. 2009, J. Math. Phys., 50, 032501
\bibitem[An \& Evans(2006)]{an06} An, J. H., \& Evans, N. W. 2006, \mnras, 369, 317
\bibitem[Blandford \& Narayan(1986)]{bla86} Blandford, R., \& Narayan, R. 1986, \apj, 310, 568
\bibitem[Chu et al.(2013)]{chu13} Chu, Z., Lin, W. P., Li, G. L., \& Kang, X. 2013, \apj, 765, 134
\bibitem[Congdon et al.(2008)]{con08} Congdon, A. B., Keeton, C. R., \& Nordgren, C. E. 2008, \mnras, 389, 398
\bibitem[Dalal(1998)]{dal98} Dalal, N. 1998, \apj, 509, L13
\bibitem[Dalal \& Rabin(2001)]{dal01} Dalal, N., \& Rabin, J. M. 2001, J. Math. Phys., 42, 1818
\bibitem[Evans \& Wilkinson(1998)]{eva98}  Evans, N. W., \& Wilkinson, M. I. 1998, \mnras, 296, 800
\bibitem[Evans \& Witt(2001)]{eva01}  Evans, N. W., \& Witt, H. J. 2001, \mnras, 327, 1260
\bibitem[Evans \& Witt(2003)]{eva03}  Evans, N. W., \& Witt, H. J. 2003, \mnras, 345, 1351
\bibitem[Finch et al.(2002)]{fin02} Finch, T. K., Carlivati, L. P., Winn, J. N., \& Schechter, P. L. 2002, \apj, 577, 51
\bibitem[Goldberg et al.(2010)]{gol10} Goldberg, D. M., Chessey, M. K., Harris, W. B., \& Richards, G. T. 2010, \apj, 715, 793
\bibitem[Hunter \& Evans(2001)]{hun01} Hunter, C., \& Evans, N. W. 2001, \apj, 554, 1227
\bibitem[Kassiola \& Kovner(1993)]{kas93} Kassiola, A., \& Kovner, I. 1993, \apj, 417, 450
\bibitem[Keeton et al.(2005)]{kee05} Keeton, C. R., Gaudi, B. S., \& Petters, A. O. 2005, \apj, 635, 35
\bibitem[Kovner(1987)]{kov87} Kovner, I. 1987, \apj, 312, 22
\bibitem[Mao \& Schneider(1998)]{mao98} Mao, S., \& Schneider, P. 1998, \mnras, 295, 587
\bibitem[Petters \& Werner(2010)]{pet10} Petters, A. O., \&  Werner, M. C., 2010, Gen. Relativ. Gravit., 42, 2011
\bibitem[Rusin \& Tegmark(2001)]{rus01} Rusin, D., \& Tegmark. M. 2001, \apj, 553, 709
\bibitem[Schneider et al.(1992)]{sch92} Schneider, P., Ehlers, J., \& Falco, E. E. 1992, Gravitational Lenses (Berlin: Springer)
\bibitem[Werner(2009)]{wer09} Werner, M. C. 2009, J. Math. Phys., 50, 082504
\bibitem[Witt \& Mao(1995)]{wit95} Witt, H. J., \& Mao, S. 1995, \apj, 447, L105
\bibitem[Witt \& Mao(2000)]{wit00} Witt, H. J., \& Mao, S. 2000, \mnras, 311, 689
\bibitem[Woldesenbet \& Williams(2012)]{wol12} Woldesenbet, A. G., \& Williams, L. L. R. 2012, \mnras, 420, 2944
\bibitem[Zakharov(1995)]{zak95} Zakharov, A. F. 1995, \aap, 293, 1

\end{thebibliography}
\end{document}